\documentclass[usenatbib]{mn2e}
\usepackage{graphicx}
\usepackage{multirow}

\title[]{How peculiar is the ``peculiar variable'' DZ Crucis (Nova Cru 2003)?}
\author[Rushton et al.]
 {M. T. Rushton$^{1,2}$, A. Evans$^2$, 
S. P. S. Eyres$^1$, 
J. Th. van Loon$^2$, 
B. Smalley$^2$ \\
$^1$Centre for Astrophysics, University of Central Lancashire, Preston, PR1
2HE, UK \\
$^2$Astrophysics Group, School of Physical \& Geographical Sciences, Keele
University, Keele, Staffordshire,  ST5 5BG, UK }

\date{Version of 22 Jan 2008}

\pagerange{\pageref{firstpage}--\pageref{lastpage}}

\def\LaTeX{L\kern-.36em\raise.3ex\hbox{a}\kern-.15em
    T\kern-.1667em\lower.7ex\hbox{E}\kern-.125emX}

\newcommand{\mic}{\mbox{$\,\mu$m}} 

\newcommand{\ltsimeq}{\raisebox{-0.6ex}{$\,\stackrel 
           {\raisebox{-.2ex}{$\textstyle <$}}{\sim}\,$}} 
\newcommand{\gtsimeq}{\raisebox{-0.6ex}{$\,\stackrel
           {\raisebox{-.2ex}{$\textstyle >$}}{\sim}\,$}}

\begin{document}
\label{firstpage}
\maketitle
\begin{abstract}
The variable star DZ~Cru was thought to be a nova when it was discovered in eruption in 2003 August. This explanation was later challenged, however, when the first spectra of the object were reported. We present near infrared spectroscopy of DZ~Cru obtained at the New Technology Telescope on 3 occasions, starting $\sim1.5$ years after outburst, with the aim of establishing the nature of the object. The spectra display H\,{\sc i}, O\,{\sc i}, [N\,{\sc i}] emission lines, together with He\,{\sc i} P~Cygni lines superposed on a dust continuum. These observations suggest the ``peculiar variable in CruxÂ'' is a classical nova.
\end{abstract}

\section{Introduction}

The ``peculiar variable'' DZ~Cru \citep{Della03} was discovered in outburst in 2003 August \citep{Tabur03}. The object was assumed to be a nova at that time, but a spectroscopic observation reported shortly after suggested other possible interpretations \citep{Della03}. Perhaps the most intriguing is the suggestion that DZ~Cru is similar to the eruptive variable V838~Mon \citep{Corradi07}. No further observations of DZ Cru have been reported, despite this possibility, and the nature of the object has remained uncertain.

V838~Mon showed an extraordinary outburst in early 2002 and has confounded all explanation and classification.
It too was assumed to be a nova when it was discovered \citep{Brown02}, but its light curve showed more than one peak and its spectra lacked many of the lines observed in novae near maximum light \citep{Crause03,Munari02}. 
Instead V838 Mon showed lines of neutral and singly ionised metals forming in a moderate velocity stellar wind \citep{Kipper04,Rushton05}. 

In the post outburst phase, V838 Mon developed strong molecular absorptions bands before stabilising as an ``L-supergiant'' \citep{Evans03a}. 
In contrast, novae show increasing levels of excitation, developing nebular, auroral, and sometimes even coronal lines \citep{Warner08}. 
Thus, an object like V838 Mon is unlikely to be mistaken for a nova during and after decline.
In this paper we present near infrared (IR) spectroscopy of DZ~Cru obtained $\sim1.5-1.9$ years post eruption, in order to establish the nature of the object.

\section{DZ Cru}
\label{ncrux}

\begin{figure*}
\centering
\includegraphics[width=7cm, angle=-90]{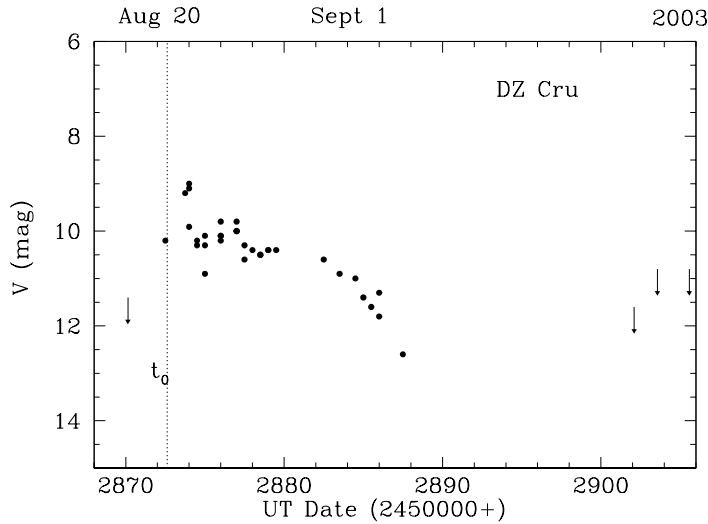}
\caption[Visual light curve of DZ~Cru]{Visual light curve of DZ~Cru based on estimates
compiled by VSNET, \citet{Tabur03}, \citet{Souza03} and
\citet{Liller03}. {\it Arrows}
denote upper limits. We assume 2003 August 20 U.T. (JD 2,451,872) to be the date of maximum light ($t_0$). The spectra reported here were obtained $\sim1.5-1.9$ years after outburst, and well outside the time interval shown.}
\label{cruxv}
\end{figure*}

On 2003 August 20 U.T., JD 2,452,872, Tabur announced the discovery of a possible
nova in Crux (hereafter DZ~Cru ${\rm R.A.}=12{\rm h}23{\rm m}16.2{\rm s},~{\rm
Dec.}=-60^{\circ}22'34''$ J2000; \citealt{Tabur03}). 
The visual light curve of DZ Cru, compiled by magnitude estimates posted on the Variable Star Network (VSNET)\footnote{http://www.kusastro.kyoto-u.ac.jp/vsnet/} 
and published in IAU circulars, is shown in
Figure~\ref{cruxv}. 
We assume maximum light $(t_0)$ occurred on 2003 August 20 U.T., JD 2,452,872, the date the outburst was first detected. The object was at least 2\,magnitudes fainter only two days previous \citep{Tabur03}.

Prediscovery images taken by Tabur, which 
date back to 2000 January, do not show an object at the position of DZ~Cru, although the outbursting star may be close
to the limiting magnitude of the CCD ($V=11.9$) on  
2003 August 18 U.T., JD 2,452,870 
(shown as an upper limit in the Figure). Palomar Sky Survey plates also show no object at this position \citep{Tabur03}. Thus the amplitude of the outburst is $\Delta V>10$, as expected for a fast, or a moderate speed, classical nova \citep{Warner08}. 

At the time of its discovery DZ~Cru had a visual magnitude of $V=10.2$. 
CCD observations taken in poor conditions a day later showed the ``possible nova'' 
at its peak observed brightness: $V=9.2$. By the following day it had become a magnitude fainter, and a further drop in brightness of $\Delta V>2$~magnitudes
began a week later, at a rate expected for a fast nova: 0.3~mag\,day$^{-1}$ \citep{Warner08}. 
On 2003 September 18 U.T., JD 2,452,901 ($t=29$ days), the visual magnitude of the object was $V>12.1$. A final, unsuccessful attempt to detect DZ~Cru was made on 2003 September 22 U.T., JD 2,452,904 ($t=32$ days), when the object was fainter than $V=11.9$.

Optical spectroscopy of DZ~Cru was obtained by \citet{Bond03} on 2003
August 21 U.T., JD 2,452,873, and \citet{Della03} on 2003 August 22 U.T., JD 2,452,874. No other
spectra of DZ~Cru have been reported hitherto. \citet{Bond03} and \citet{Della03} reported 
H$\alpha$ and H$\beta$ P~Cygni lines, with weak emission
components. According to
\citet{Della03} however, the spectrum
showed few other signatures of a classical nova. The
expansion velocity was found to be  
$\sim500$\,km\,s$^{-1}$ from the P~Cygni lines, a value not expected for a fast nova ($\sim800-2500$\,km\,s$^{-1}$; \citealt{Warner08}). 
DZ~Cru also appeared too red to
be a nova. The $B$ band magnitude on
2002 August 22 U.T., JD 2,452,874, is $B=11.16\pm0.05$ \citep{Liller03}, giving a $B-V$ index of $+1.25\pm0.05$; if DZ Cru is a nova one would expect $(B-V)=+0.23\pm0.05$ at $t_0$ \citep{vandenBergh87}. These peculiarities led to speculation the ``peculiar variable in Crux'' may be a ``V838 Mon-type object", or may be experiencing a very late thermal pulse (VLTP) \citep{Della03}. 

\section{Observations}
\label{obs}

Near-IR spectroscopy of DZ~Cru was obtained on three occasions at ESO's New Technology Telescope (NTT), La Silla, in a five month period starting $\sim1.5$ years after outburst. An observing log is shown in Table~\ref{log}. The observations employed the Son OF Isaac (SOFI) infrared spectrograph and imaging camera \citep{Lidman02}, with the blue and red low resolution grisms, and a $0.6''$ slit. The wavelength coverage on each occasion is $0.95-2.52\,\mu$m. The resolving power is $R\sim1000$. 

The data were obtained in nod-on-slit mode, and sky emission lines were removed by subtracting off-source spectra from on-source spectra.
Atmospheric absorption was eliminated to a large extent by dividing the sky-subtracted spectra by spectra of the standard stars listed in Table~\ref{log}. In order to maximise cancellation of telluric lines, atmospheric absorption features in the calibration stars were used to realign the wavelength axis of the target spectra prior to division. The hydrogen absorption lines in the standards were manually snipped out of the data to avoid spurious emission in the ratioed spectra.    
Flux calibration was accomplished by multiplying the ratioed spectra by the spectrum of a blackbody with appropriate effective temperature for the calibration star. The uncertainties in the fluxes are $\sim~10-20\%$.  
Wavelength calibration was achieved from Xenon arc spectra, and is accurate to $\pm0.003\,\mu$m ($1\sigma$). 
In spectral regions with good transmission, the signal-to-noise ratio of the data is $\sim200$ in the first two observations, and $\sim50$ in the final observation.

\begin{table*}
\caption{Observing log.}
\begin{tabular}{lccccccccc}
\hline

\multicolumn{1}{l}{UT Date} &
\multicolumn{1}{c}{JD$^a$} &
\multicolumn{1}{c}{Age$^b$} &
\multicolumn{1}{c}{Exposure} &
\multicolumn{1}{c}{Flux std} &
\multicolumn{1}{c}{Airmass$^c$} &
\multicolumn{1}{c}{Airmass$^c$} &
\multicolumn{1}{c}{$J^d$} &
\multicolumn{1}{c}{$H^d$} &
\multicolumn{1}{c}{$K^d$} \\

\multicolumn{1}{l}{} &
\multicolumn{1}{l}{} &
\multicolumn{1}{c}{(days)} &
\multicolumn{1}{c}{(s)} &
\multicolumn{1}{c}{} & 
\multicolumn{1}{c}{(target)} & 
\multicolumn{1}{c}{(standard)} & 
\multicolumn{1}{c}{} &
\multicolumn{1}{c}{} \\

\hline

\multicolumn{1}{l}{2005 February 16} &
\multicolumn{1}{c}{3417} &
\multicolumn{1}{c}{544} &
\multicolumn{1}{c}{151} &
\multicolumn{1}{c}{HIP037732 (G1 V)}&
\multicolumn{1}{c}{1.23} & 
\multicolumn{1}{c}{1.19} & 
\multicolumn{1}{c}{13.8}&
\multicolumn{1}{c}{11.5} &
\multicolumn{1}{c}{9.7} \\

\multicolumn{1}{l}{2005 June 16} &
\multicolumn{1}{c}{3538} &
\multicolumn{1}{c}{665} &
\multicolumn{1}{c}{480} &
\multicolumn{1}{c}{HIP062871 (G0 V)}&
\multicolumn{1}{c}{1.17} &
\multicolumn{1}{c}{1.18} &
\multicolumn{1}{c}{14.1} &
\multicolumn{1}{c}{12.2} &
\multicolumn{1}{c}{10.4} \\

\multicolumn{1}{l}{2005 July 22} &
\multicolumn{1}{c}{3574} &
\multicolumn{1}{c}{701} &
\multicolumn{1}{c}{480} &
\multicolumn{1}{c}{HIP92233 (F8 V)}&
\multicolumn{1}{c}{1.59} & 
\multicolumn{1}{c}{1.56} & 
\multicolumn{1}{c}{14.2}&
\multicolumn{1}{c}{12.5}&
\multicolumn{1}{c}{11.0} \\

\hline

\multicolumn{5}{l}{All observations were conducted at the NTT using SOFI and a $0.6''$ slit.}\\ 
\multicolumn{5}{l}{Resolving power is $R\sim1000$ on each occasion.}\\
\multicolumn{5}{l}{$^a$Julian date--2,450,000.}\\
\multicolumn{5}{l}{$^b$2003 August 20 U.T., JD 2,452,872 is taken as $t_0$.}\\
\multicolumn{5}{l}{$^c$Airmass near mid-point of each set of observations.}\\
\multicolumn{8}{l}{$^d$Approximate magnitude. Deduced by convolving the spectra with the relevant band profile.}

\end{tabular}
\label{log}
\end{table*}

\section{Results}
\label{overspec}

\begin{figure}
\centering
\includegraphics[width=8cm, angle=0]{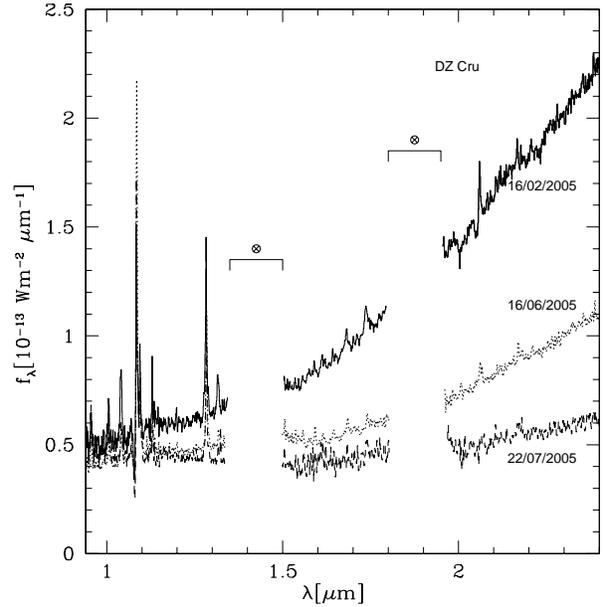}
\caption[]{NTT time series spectra of DZ~Cru in the $JHK$ bands ($0.95-2.52\,\mu$m), spanning from 2005 February 16 to 2005 July 22 U.T ($\sim1.5-2$ years from maximum brightness). Gaps in the data in the $\sim1.34-1.50\,\mu$m and $\sim1.80-1.95\,\mu$m regions are owing to strong absorption from terrestrial H$_2$O. Identifications of the spectral lines are given in Figure~\ref{cruxspec}. Date format is DD/MM/YY.}
\label{evolution}
\end{figure}

\begin{figure}
\centering
\includegraphics[width=8cm, angle=0]{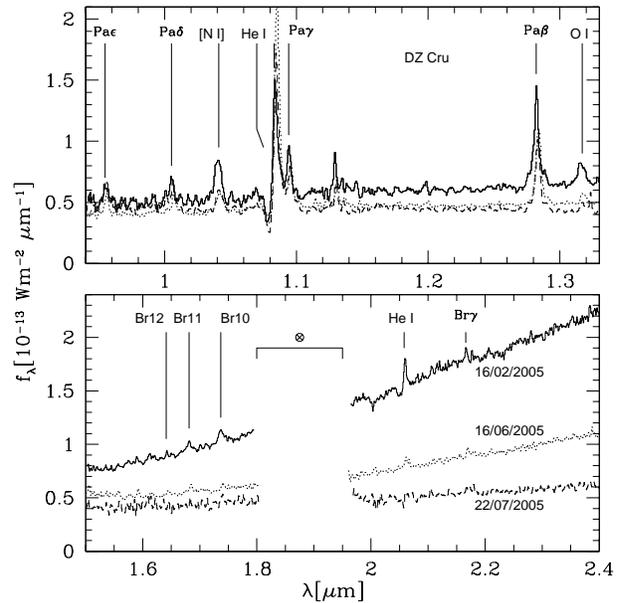}
\caption[Near-IR spectrum of DZ~Cru on 2005 February 16]{Enlargement of the near-IR spectra 
of DZ~Cru shown in Figure~\ref{evolution}, with the individual spectral features identified. Br10, Br11, and Br12 are hydrogen Brackett lines labelled with their upper level quantum numbers. Date format is DD/MM/YY.}
\label{cruxspec}
\end{figure}

Figure~\ref{evolution} shows the evolution of the near-IR spectrum of DZ~Cru between 2005 February 16 U.T., JD 2,453,417, and 2005 July 22 U.T., JD 2,453,574 ($t\sim544$ days to $t\sim701$ days).   
Data are not shown in the $\sim1.34-1.50\,\mu$m and $\sim1.80-1.95\,\mu$m ranges, owing to poor telluric cancellation in those regions. Figure~\ref{cruxspec} displays the same spectra expanded in the blue ($\sim0.95-1.33\mic$) and red ($\sim1.56-2.40\mic$) regions to show spectral lines clearly. Approximate $JHK$ magnitudes of DZ Cru, derived by convolving the spectra with the relevant band profile, are given in Table~\ref{log}. 

The near-IR spectrum of DZ~Cru shows an emission line spectrum superposed on a dust continuum on all dates of observation.
Table~\ref{linelist} lists observed wavelengths, proposed identifications, full width at half maximum (FWHM),
and integrated fluxes
of most spectral lines in the data.
The spectra contain resolved Pa$\beta$,
$\gamma$, $\delta$, $\epsilon$ 
and weak Br$\gamma$ emissions, 
along with higher order ($n\gtsimeq12$) hydrogen
Brackett lines. 
Blends of He\,{\sc i} $^3$S--$^3$P$^0$ lines at
1.08321, 1.08332 and 1.08333\,$\mu$m
\citep{vanhoof06}
give rise to the most prominent spectral line, which clearly displays a P~Cygni-type profile. The identification 
with helium is secured by
the presence (on 2005 February 16 at least) of the $^1$S--$^1$P$^0$ He\,{\sc i} singlet at $2.058\,\mu$m, the only
other spectral line with a P~Cygni
profile. The forbidden
[N\,{\sc i}] $^2\rm D^0-^2P^0$ doublet at 
1.040\,$\mu$m and the allowed O\,{\sc i} $^3$P$-^3$S$^0$ triplet at 
1.316\,$\mu$m are likely identifications of the remaining features.

\begin{figure}
\centering
\includegraphics[width=8cm, angle=0]{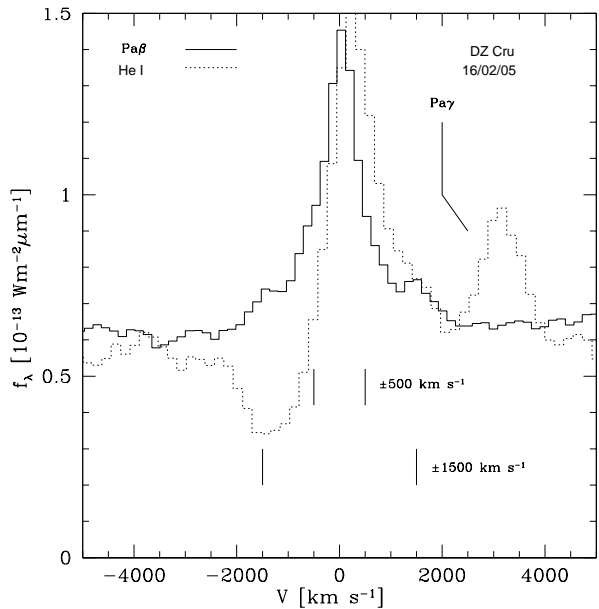}
\caption[The Pa$\beta$ line in DZ~Cru]{Close up of the Pa$\beta$ line in DZ~Cru on 2005 February 16 ({\it solid}), showing the line structure, which is similar on all dates of observation. The helium line at $1.083\,\mu$m from the same spectrum is shown for comparison
({\it dotted}). The emission feature redward of the He\,{\sc i} line is
Pa$\gamma$.}
\label{pbeta}
\end{figure}

The mean expansion velocities deduced from deconvolved FWHM of the H\,{\sc i} lines are constant over the course of observation; they are
$527\pm109$\,km\,s$^{-1}$ (2005 February 15), $496\pm50$\,km\,s$^{-1}$ (2005 June 16), and $462\pm35$\,km\,s$^{-1}$ (2005 July 22), in agreement with outflow velocities reported by \cite{Della03} near $t_0$. The expansion velocity determined from Br$\gamma$ is lower on all dates (Table~\ref{linelist}). This line is swapped by the thermal continuum, and was not used in calculating the mean. The He\,{\sc i}  P~Cygni lines indicate mean expansion velocities of $\sim1500$\,km\,s$^{-1}$; weak emission components are seen in H\,{\sc i} at this velocity (see Figure~\ref{pbeta}) - similar line profiles were observed in the nova V2487 Oph \citep{Lynch00}. 

The most noticeable changes that occurred over the course of our observations are an increase in the strength of the He\,{\sc i} $1.083\,\mu$m emission component, and a sharp decrease in the $K$-band flux.
The evolution of the He\,{\sc i} $1.083\,\mu$m line is similar to its development in nova PW Vul \citep{Williams96}. \cite{Williams96} attributed the behaviour to the hardening radiation field from the central source and its effect on the He\,{\sc i} 2$^3$S--$2^3$P collisional cross section. For electron temperatures in the range of $T_{\rm e}=5000-20,000$\,K, the cross section varies as $T_{\rm e}^{1.26}$.

 The decline in the $K$-band flux is presumably owing to dispersal and cooling of the dust over the $\sim157$ day interval of observation.
We find the dust temperatures $T_{\rm d}$ by fitting the dust continuum with $B(\lambda, T_{\rm d})\lambda^{-\beta}$, where $B$ is the Planck function at temperature $T_{\rm d}$ and $\beta$ is the emissivity index, which depends on the nature of the dust; for amorphous carbon $\beta\simeq1$ \citep{Mennella98}. 
We find dust temperatures of $T_{\rm d}=690\pm40$\,K (February) and $T_{\rm d}=620\pm50$\,K (June), where the uncertainty is owing to the limited wavelength range of the data. 

\section{The nature of DZ~Cru}
\label{nature}

There are a number of possible explanations for the outburst of DZ~Cru. We address each in turn.

\subsection{A V838 Mon type event?}

This is referring to the eruption of V838 Mon in early 2002. The object is perhaps best known for a spectacular light echo that developed around the star shortly after its discovery \citep{Bond03a}. It is thought V838 Mon belongs to a new class of eruptive variable, along with M31~RV \citep{Rich89}, V4332 Sgr \citep{Martini99}, and M85 OT2006-1 \citep{Rau07}, although the latter has been interpreted as a low luminosity type-II plateau supernova \citep{Pastorello07}.

Our near-IR spectra of DZ~Cru bear no resemblance whatsoever to those of V838 Mon $\sim1.5-2$ years after its outburst, nor to those observed at any stage in the evolution of the object. By that time V838 Mon had stabilised somewhat as a very cool supergiant, with a near-IR spectrum dominated by strong molecular bands of H$_2$O, CO, and metal oxides \citep{Evans03a}.
V838 Mon did not show high excitation He\,{\sc i} lines, nor did it display outflow velocities greater than $\sim500$\,km\,s$^{-1}$ \citep{Kipper04}.
Thus we conclude that the outburst of DZ~Cru  is unlikely to have been a V838 Mon type event. 

\subsection{A born-again AGB?}

The central stars of planetary nebulae may experience a VLTP, resulting in a dramatic brightening of the star. They are known as ``born-again giants'' because they
retrace their evolution toward the region of the HR diagram occupied by the AGB stars and repeat their post AGB evolution. 
The eruptions of V605~Aql in 1919 \citep{Lechner04}, and Sakurai's Object in 1996 \citep{Evans02a} have been attributed to VLTPs.

The photometric behaviour of DZ~Cru is unlike that of Sakurai's Object, the best studied born-again giant:
the amplitude of the outburst of DZ~Cru is at least $\Delta V\simeq5$ magnitudes larger, and the time it took to attain maximum light is much shorter:  $\ltsimeq2$~days as opposed to $\sim1$~year. 
The light curve of Sakurai's Object has been interpreted in terms of a slowly expanding photosphere \citep{Duerbeck02}. Its spectral evolution suggested a rapidly cooling star, and deep molecular absorption bands developed shortly after maximum light \citep{Eyres99}. Unlike DZ Cru, no emission lines, or P Cygni profiles were observed after more than 2 years since outburst. 

During a VLTP convection in the region surrounding the helium shell leads to mixing and changes in the chemical abundances.
This process renders
born-again giants hydrogen-deficient and C-rich \citep{Herwig05}.
Consequently, they
display weaker H\,{\sc i} lines than stars with similar effective temperature.  
Thus, strong hydrogen lines, such as those present in the spectrum of DZ~Cru, are not expected in a 
born-again giant. We conclude this explanation is unlikely.

\subsection{A nova type object?}

The spectra presented here leave us in no doubt DZ~Cru is a nova, for H\,{\sc i}, He\,{\sc i}, [N\,{\sc i}], and O\,{\sc i} lines appear in their near-IR spectra (e.g. \citealt{Lynch06,Venturini04,Lynch04,Rudy03,Evans03}), and extensive dust shells are known to have formed around some novae \citep{Evans08}. Furthermore, 
the outflow velocities in DZ~Cru are within the range shown by novae \citep{Warner08}.

Principal ejection velocities in novae range from $\sim500$\,km\,s$^{-1}$ to $\sim2500$\,km\,s$^{-1}$ depending on speed class. For DZ Cru this velocity is $\sim500$\,km\,s$^{-1}$ \citep{Della03}. As we noted above, the outburst amplitude and decline rate suggest a fast nova, and the principal velocities of such objects are $\sim800-2500$\,km\,s$^{-1}$ \citep{Warner08}. This discrepancy suggests the object is deviating from the expansion velocity-rate of decline relationship of classical novae. However, DZ~Cru is not alone in showing this behaviour: QU~Vul \citep{Rosino92}, V838~Her \citep{Harrison94}, and V2487~Oph \citep{Lynch00} are just three other examples.
 
DZ~Cru shows two outflows in our data: one at the principal velocity and a second at $\sim1500$\,km\,s$^{-1}$. 
The fast outflow -- unreported near $t_0$ -- 
is likely to have emerged in the decline phase, when multiple velocity components appear in the spectral lines of novae \citep{Warner08}.
A similar situation was observed in V723~Cas several hundred days after $t_0$, when this slow nova showed hydrogen emission lines indicating an outflow velocity of $\sim300$\,km\,s$^{-1}$ and He\,{\sc i} P~Cygni lines indicating an outflow velocity of $\sim1500$\,km\,s$^{-1}$ \citep{Evans03}. 
There are some major differences between DZ~Cru and V723~Cas, however: whereas DZ~Cru was dusty and non-coronal, V723~Cas was dust-free and strongly-coronal. The absence of dust in V723 Cas could be related to its coronal behaviour. Novae that show a coronal phase are generally poor dust producers \citep{Evans08}. 

The only unresolved problem is the colour of DZ~Cru on $t=2$ days  [$(B-V)=+1.25\pm0.05$], which is $1.02\pm0.08$ magnitudes redder than intrinsic colours of classical novae at $t_0$ \citep{vandenBergh87}.
The colour of DZ~Cru is affected by an unknown amount of interstellar reddening, however, and we need to account for this before making a comparison with the intrinsic colours of novae. According to the relevant extinction curve in \cite{Marshall06}, the reddening rises rapidly with distance along the line of sight towards DZ~Cru (see $\S$\ref{dist}). It is probable the unusual colour near $t_0$ is owing to heavy interstellar reddening in this direction; if so, the distance to DZ Cru is $d\gtsimeq7$\,kpc. 

\begin{table*}
\caption{DZ~Cru linelist. Date format is DD/MM/YY.}
\begin{tabular}{lccccccc}
\hline

\multicolumn{1}{l}{Observed $\lambda^a$} &
\multicolumn{1}{c}{Identification} &
\multicolumn{3}{c}{FWHM$^{b,c}$} &
\multicolumn{3}{c}{Flux$^b$} \\ [4pt]

\multicolumn{1}{c}{($\mu$m)} &
\multicolumn{1}{c}{} &
\multicolumn{3}{c}{(km\,s$^{-1}$)} &
\multicolumn{3}{c}{(10$^{-16}$Wm$^{-2}$)} \\ [4pt]

\multicolumn{1}{c}{16/02/05} &
\multicolumn{1}{c}{} &
\multicolumn{1}{c}{16/02/05} &
\multicolumn{1}{c}{16/06/05} &
\multicolumn{1}{c}{22/07/05} &
\multicolumn{1}{c}{16/02/05} &
\multicolumn{1}{c}{16/06/05} &
\multicolumn{1}{c}{22/07/05} \\

\hline

\multicolumn{1}{l}{0.9554} &
\multicolumn{1}{l}{H\,{\sc i} Pa$\epsilon$ 0.9549} &
\multicolumn{1}{c}{990} &
\multicolumn{1}{c}{1048} &
\multicolumn{1}{c}{1011} &
\multicolumn{1}{c}{$0.49\pm0.04$} & 
\multicolumn{1}{c}{$0.60\pm0.03$} & 
\multicolumn{1}{c}{$0.46\pm0.02$} \\

\multicolumn{1}{l}{1.0054} &
\multicolumn{1}{l}{H\,{\sc i} Pa$\delta$ 1.0052} &
\multicolumn{1}{c}{872} &
\multicolumn{1}{c}{1095} &
\multicolumn{1}{c}{831} &
\multicolumn{1}{c}{$0.53\pm0.04$} & 
\multicolumn{1}{c}{$0.38\pm0.02$} & 
\multicolumn{1}{c}{$0.92\pm0.06$} \\

\multicolumn{1}{l}{1.0405} &
\multicolumn{1}{l}{[N\,{\sc i}] 1.0401} &
\multicolumn{1}{c}{1568} &
\multicolumn{1}{c}{1738} &
\multicolumn{1}{c}{$-$} &
\multicolumn{1}{c}{$1.72\pm0.09$} & 
\multicolumn{1}{c}{$0.92\pm0.03$} & 
\multicolumn{1}{c}{$-$} \\

\multicolumn{1}{l}{$-^{d}$} &
\multicolumn{1}{l}{He\,{\sc i} 1.0833} &
\multicolumn{1}{c}{$-$} & 
\multicolumn{1}{c}{$-$} & 
\multicolumn{1}{c}{$-$} & 
\multicolumn{1}{c}{$-$} & 
\multicolumn{1}{c}{$-$} & 
\multicolumn{1}{c}{$-$} \\

\multirow{1}{*}{1.0944} &
\multirow{1}{*}{H\,{\sc i} Pa$\gamma$ 1.0941} &
\multirow{1}{*}{885} &
\multirow{1}{*}{1048} &
\multirow{1}{*}{889} &
\multirow{1}{*}{$1.43\pm0.04$} & 
\multirow{1}{*}{$1.51\pm0.03$} & 
\multirow{1}{*}{$1.56\pm0.05$} \\

\multirow{1}{*}{1.2824} &
\multirow{1}{*}{H\,{\sc i} Pa$\beta$ 1.2822} &
\multirow{1}{*}{1008} &
\multirow{1}{*}{1055} &
\multirow{1}{*}{961} &
\multirow{1}{*}{$2.45\pm0.21$} & 
\multirow{1}{*}{$2.29\pm0.33$} & 
\multirow{1}{*}{$2.13\pm0.24$} \\

\multicolumn{1}{l}{1.3163} &
\multicolumn{1}{l}{O\,{\sc i} 1.3168} &
\multicolumn{1}{c}{1437} &
\multicolumn{1}{c}{1042} &
\multicolumn{1}{c}{$-$} &
\multicolumn{1}{c}{$0.90\pm0.05$} & 
\multicolumn{1}{c}{$0.40\pm0.02$} & 
\multicolumn{1}{c}{$-$} \\

\multicolumn{1}{l}{1.6811} &
\multicolumn{1}{l}{H\,{\sc i} Br11 1.6811} &
\multicolumn{1}{c}{1047} &
\multicolumn{1}{c}{836} &
\multicolumn{1}{c}{$-$} &
\multicolumn{1}{c}{$0.51\pm0.04$} & 
\multicolumn{1}{c}{$0.40\pm0.02$} & 
\multicolumn{1}{c}{$-$} \\

\multicolumn{1}{l}{1.7369} &
\multicolumn{1}{l}{H\,{\sc i} Br10 1.7367} &
\multicolumn{1}{c}{1519} &
\multicolumn{1}{c}{869} &
\multicolumn{1}{c}{$-$} &
\multicolumn{1}{c}{$0.98\pm0.03$} & 
\multicolumn{1}{c}{$0.33\pm0.02$} & 
\multicolumn{1}{c}{$-$} \\

\multicolumn{1}{l}{$-^{d}$} &
\multicolumn{1}{l}{He\,{\sc i} 2.0587} &
\multicolumn{1}{c}{$-$} & 
\multicolumn{1}{c}{$-$} & 
\multicolumn{1}{c}{$-$} & 
\multicolumn{1}{c}{$-$} & 
\multicolumn{1}{c}{$-$} & 
\multicolumn{1}{c}{$-$} \\

\multicolumn{1}{l}{2.1667} &
\multicolumn{1}{l}{H\,{\sc i} Br$\gamma$ 2.1661} &
\multicolumn{1}{c}{477} &
\multicolumn{1}{c}{554} &
\multicolumn{1}{c}{$-$} &
\multicolumn{1}{c}{$0.47\pm0.04$} & 
\multicolumn{1}{c}{$0.46\pm0.02$} & 
\multicolumn{1}{c}{$-$} \\

\hline
\multicolumn{6}{l}{$^{a}$For the spectrum obtained on 2005 February 16. No significant shift had occurred by the later dates.} \\
\multicolumn{6}{l}{$^{b}$Deconvolved for instrumental profile and deduced from gaussian fitting.} \\
\multicolumn{6}{l}{$^{c}$Uncertainty $\sim\pm150$\,km\,s$^{-1}$.} \\
\multicolumn{6}{l}{$^{d}$P~Cygni-type profile.} \\

\end{tabular}
\label{linelist}
\end{table*}

\section{Distance}
\label{dist}

Absolute magnitudes of classical novae can be found from the magnitude at maximum ($M_{\rm V}^{\rm max}$)-rate of decline (MMRD) relationship which states $M_{\rm V}^{\max}$ is linearly proportional to the logarithm of the time taken $t_2$ to decline by $\Delta V=2$ magnitudes from $t_0$ \citep{Warner08}. 
From the light curve (Figure~\ref{cruxv}) we find $t_2=15$ days. 
We note there is a spike in the light curve on $t=1$ days; this could be owing to a flare, which are thought to arise from wind interactions, or variations in the mass loss, and are common phenomena in slow novae, e.g. HR Del \citep{Drechsel77}, V1548 Aql \citep{Kato01} and V723 Cas \citep{Munari96}.
The most recent calibration of the $M^{\rm max}_{\rm V}-t_2$ relation is:
$M^{\rm max}_{\rm V}=(-11.32\pm0.44)+(2.55\pm0.32)\log(t_2)$   
\citep{Downes00}, where $t_2$ is in days. We obtain $M^{\rm max}_{\rm V}=-8.3\pm0.5$ magnitudes, which is bright for a nova with a principal velocity of $\sim500$\,km\,s$^{-1}$, as this velocity implies a longer decline time.

In order to determine the distance to the nova from this result we need the interstellar extinction along the line of sight in the visual.  
The galactic coordinates of DZ~Cru are $l=299.45^{\circ}$, $b=+2.31^{\circ}$. According to dust maps in \cite{Schlegel98}, the total extinction along this line of sight is $A_{\rm V}=3.6$, which agrees with results in \cite{Marshall06}, showing that $A_{\rm V}$ rises linearly beyond $d=3$\,kpc, reaching $A_{\rm V}=3.4\pm0.3$ at $d=8.02\pm1.33$\,kpc. Assuming this value for the extinction, we obtain a distance to DZ~Cru of $d=10.5\pm2.8$\,kpc with $M^{\rm max}_{\rm V}=-8.3\pm0.5$ magnitudes. A similarly large distance is found from the relation which states all novae have similar absolute magnitudes ($-6.05\pm0.44$\,magnitudes) near $t\sim15$ days \citep{Downes00}: $d=11.1\pm2.8$\,kpc. 

In view of the discrepancy between $t_2$ and the principal velocity, the distance may be poorly determined from the light curve. 
As an alternative, we estimate the distance to DZ~Cru from the $t_2$ implied by its principal velocity, while being mindful of the lower extinction for distances of $d<8$\,kpc in the direction of the object. 
Fitting the \cite{Marshall06} extinction curve towards DZ~Cru with a straight line using the {\sc Fitexy} routine \citep{Press92}, we find $A_{\rm V}$ increases by $0.40\pm0.03$ mags kpc$^{-1}$. 
From the expansion velocity-rate of decline relationship of classical novae \citep{Warner08}, we find $t_2=55$\,days for a principal velocity of 500\,km\,s$^{-1}$. This result leads to $M_{\rm V}^{\rm max}=-6.9\pm0.5$ magnitudes from the MMRD. Assuming the extinction gradient above, the distance is then $d=7.3$\,kpc. This nearer estimate is still consistent with the lower limit implied by the reddening. Thus we have possible distances of $d\sim7-12$\,kpc. As a compromise, we adopt $d=9$\,kpc, and assume this distance throughout the remainder of this paper.

\section{Discussion}

Now the nature and distance are determined we are in a position to estimate the dust mass $M_{\rm d}$. If we assume the dust shell is optically thin in the IR, and dominated by amorphous carbon \citep{Mennella98}, 
then $M_{\rm d}$ (in $M_{\odot}$) can be found from 

\begin{eqnarray}
M_{\rm d}\simeq6.8\times10^{11}f_{\lambda}d_{\rm kpc}^2\lambda/B(\lambda, T_{\rm d}), \nonumber
\end{eqnarray}
where $f_{\lambda}$ is the flux density in Wm$^{-2}\mu$m$^{-1}$, $\lambda$ is wavelength in $\mu$m, and 
$d_{\rm kpc}$ is the distance in kpc. We use the $K$-band magnitude to deduce $f_{\lambda}$ (Table~\ref{log}), as the dust emission is more important at longer wavelengths. We find $M_{\rm d}=3\times10^{-8}\,M_{\odot}$ (February) and $5\times10^{-8}\,M_{\odot}$ (June). 
For a canonical gas-to-dust ratio of 100, the total mass ejected is $M_{\rm g}\sim10^{-6}\,M_{\odot}$, although this mass is uncertain, owing to uncertainty in the gas-to-dust ratio. In novae, ratios from $10^{1}$ to $10^4$ have been reported \citep{Gehrz98}.  

If we know the electron temperature $T_e$ and density $N_e$ of the gas, we can calculate its mass from the H\,{\sc i} line fluxes. 
Since the luminosity of Pa$\beta$ is  
$L_{\rm Pa\beta}=h\nu\alpha_{\rm Pa\beta}(N_e, T_e)N_e^2V$, $N_e$, in cm$^{-3}$, can be found from 
\begin{eqnarray}
N_e\simeq2.0\times10^{21}d_{\rm kpc}\left(\frac{F_{\rm Pa\beta}T_{\rm e}}{f(vt)^3}\right)^\frac{1}{2},
\nonumber
\end{eqnarray}
where $F_{\rm Pa\beta}$ is the line flux in Wm$^{-2}$, $T_{\rm e}$ is in $10^4$\,k, $f$ is the filling factor, $v$ is the velocity of the ejecta in km\,s$^{-1}$, $t$ is the time since outburst in days, and $V$ is the volume of the emitting region [$=f\frac{4}{3}\pi (vt)^3$].
For the recombination coefficient $\alpha(N_e, T_e)$, we have used the expression
\begin{eqnarray}
\alpha_{\rm Pa\beta}(N_e, T_e)\simeq\frac{1.2\times10^{-14}}{T_{\rm e}}\,{\rm cm}^{3}\,{\rm s}^{-1}, \nonumber
\end{eqnarray}
which is an analytical fit to data for case B in \cite{Hummer87}. Their data cover $T_e=3\times10^3-3\times10^4$K, and $N_e=10^2-10^{10}$\,cm$^{-3}$. This expression gives a reasonble approximation of all $\alpha_{\rm Pa\beta}(N_e, T_e)$ 
in \cite{Hummer87}.
For $T_e=10^4$\,K, $N_e=2\times10^6\,f^{-1/2}$\,cm$^{-3}$. 

Since $M_{\rm g}=N_e\mu m_{\rm H}V$, the mass of the ejected shell, in $M_{\odot}$, can be found from 
\begin{eqnarray}
M_{\rm g}\simeq8.1\times10^{15}\frac{\mu d_{\rm kpc}^2F_{\rm Pa\beta}T_{\rm e}}{N_e},
\nonumber
\end{eqnarray}
where $\mu$ is the mean molecular weight, and $m_{\rm H}$ is the mass of a hydrogen atom. Assuming $\mu=1.5$, we deduce a gas mass of $M_{\rm g}=1\times10^{-4}\,f^{1/2}\,M_{\odot}$. In novae filling factors as low as $f\sim10^{-4}$ have been calculated \citep{Gehrz98}. Thus, we derive $M_{\rm g}=10^{-6}-10^{-4}\,M_{\odot}$. 
This result is consistent with the masses ejected by novae, and further supports our interpretation of DZ~Cru.

\section{Conclusions}
\label{summ}

In this paper we presented near-IR spectra of DZ~Cru obtained $\sim1.5-1.9$ years post eruption.
This star was thought to be a nova when it was discovered in outburst in 2003 August, but optical spectroscopy reported a few days later suggested otherwise. Since then the ``peculiar variable'' has been thought to be a V838 Mon-type object, or a born-again giant.
Our spectra however, show beyond all doubt DZ Cru is a classical nova. They display
 many of the usual signatures of classical novae (H\,{\sc i}, O\,{\sc i}, and [N\,{\sc i}] emission lines) and He\,{\sc i}
P~Cygni lines, indicating outflow speeds of $\sim1500$\,km\,s$^{-1}$.  Further, the level of the continuum increases with increasing wavelength, owing to emission by dust.
As expected for a classical nova, the total mass ejected in the eruption is $\sim10^{-6}-10^{-4}\,M_{\odot}$.


\begin{thebibliography}{}

\bibitem[\protect\citeauthoryear{Bond}{2003}]{Bond03} Bond, H. E. 2003, IAUC, 8185

\bibitem[\protect\citeauthoryear{Bond et al.}{2003}]{Bond03a} Bond, H. E., Henden, A., Levay, Z. G., Panagia, N., Sparks, W. B., Starrfield, S., Wagner, R. M., Corradi, R. L. M., Munari, U. 2003, Nat, 422, 405

 \bibitem[\protect\citeauthoryear{Brown et al.}{2002}]{Brown02} Brown, N. J.,  Waagen, E. O., Scovil, C., Nelson, P., Oksanen, A., Solonen, J., Price, A. 2002, IAUC 7785

\bibitem[\protect\citeauthoryear{Corradi \& Munari}{2007}]{Corradi07} Corradi, R. L. M., Munari, U. 2007, ed. The Nature of V838 Mon and its Light Echo, ASP Conf. Ser., 363 

\bibitem[\protect\citeauthoryear{Crause et al.}{2003}]{Crause03} Crause, L. A., Lawson, W. A., Menzies, J. W., Marang, F. 2003, MNRAS, 341, 785

\bibitem[\protect\citeauthoryear{Della Valle et al.}{2003}]{Della03} Della Valle, M., Hutsemekers, D., Savianne, T., Wenderoth, E. 2003, IAUC, 8185

\bibitem[\protect\citeauthoryear{Downes \& Duerbeck}{2000}]{Downes00} Downes, R. A., Duerbeck, H. W. 2000, AJ, 120, 2007

\bibitem[\protect\citeauthoryear{Drechsel et al.}{1977}]{Drechsel77} Drechsel, H., Rahe, J., Duerbeck, H. W., Kohoutek, H. W., Seitter, W. C. 1977, A\&AS, 30, 323 

\bibitem[\protect\citeauthoryear{Duerbeck}{2002}]{Duerbeck02} Duerbeck, H. W. 2002, Ap\&SS, 279, 5

\bibitem[\protect\citeauthoryear{Evans \& Smalley}{2002}]{Evans02a} Evans, A., Smalley, B., 2002, Ap\&SS, 279 

\bibitem[\protect\citeauthoryear{Evans et al.}{2003a}]{Evans03a} Evans, A., Geballe, T. R., Rushton, M. T., Smalley, B., van Loon, J. Th., Eyres, S. P. S., Tyne, V. H. 2003a, MNRAS, 343, 1054 

\bibitem[\protect\citeauthoryear{Evans et al.}{2003b}]{Evans03} Evans, A., Gehrz, R. D., Geballe, T. R., Woodward, C. E., Salama, A., Antolin Sanchez, R., Starrfield,
S. G., Krautter, J., Barlow, M., Lyke, J. E., Hayward, T.L., Eyres, S. P. S., Greenhouse, M. A., Hjellming, R. M., 
Wagner, R. M., Pequignot, D. 2003b, AJ, 126, 1981

\bibitem[\protect\citeauthoryear{Evans \& Rawlings}{2008}]{Evans08} Evans, A., Rawlings, J. M. C. 2008, in Classical Novae, eds. M. F. Bode \& A. Evans, 2nd edition, Cambridge University Press, in press

\bibitem[\protect\citeauthoryear{Eyres et al.}{1999}]{Eyres99} Eyres, S. P. S., Smalley, B., Geballe, T. R., Evans, A., Asplund, M., Tyne, V. H. 1999, MNRAS, 307, L11

\bibitem[\protect\citeauthoryear{Gehrz et al.}{1998}]{Gehrz98} Gehrz, R. D., Truran J. W., Williams, R. E., Starrfield, S. 1998, PASP, 110, 3

\bibitem[\protect\citeauthoryear{Harrison \& Stringfellow}{1994}]{Harrison94} Harrison, T. E., Stringfellow, G. S. 1994, ApJ, 437, 827

\bibitem[\protect\citeauthoryear{Herwig}{2005}]{Herwig05} Herwig, F. 2005, ARA\&A, 43 435 

\bibitem[\protect\citeauthoryear{Hummer \& Storey}{1987}]{Hummer87} Hummer, D. G., Storey, P. J. 1987, MNRAS, 224, 801 

\bibitem[\protect\citeauthoryear{Kato \& Takamizawa}{2001}]{Kato01} Kato, T., Takamizawa, K. 2001, IBVS, 5100 

\bibitem[\protect\citeauthoryear{Kipper et al.}{2004}]{Kipper04} Kipper, T., Klochkova, V. G., Annuk, K., Hirv, A., Kolka, I., Leedj\"{a}rv, L., Puss, A., Skoda, P., Slechta, M. 2004, A\&A, 416, 1107 

\bibitem[\protect\citeauthoryear{Lechner \& Kimeswenger}{2004}]{Lechner04} Lechner, M. F. M., Kimeswenger, S. 2004, A\&A, 426, 145

\bibitem[\protect\citeauthoryear{Lidman \& Cuby}{2002}]{Lidman02} Lidman, C., Cuby, J.-G. 2002, SOFI User's Manual, Issue 1.4 

\bibitem[\protect\citeauthoryear{Liller et al.}{2003}]{Liller03} Liller, W., Shida, R. Y., Aguiar, J. G. S., Pearce, A. 2003, IAUC, 8188

\bibitem[\protect\citeauthoryear{Lynch et al.}{2000}]{Lynch00} Lynch, D. K., Rudy, R. J., Mazuk, S., Puetter, R. C. 2000, ApJ, 541, 791 

\bibitem[\protect\citeauthoryear{Lynch et al.}{2004}]{Lynch04} Lynch, D. K., Wilson, J. C., Rudy, R. J., Venturini, C., Mazuk, S., Miller, N. A., Puetter, R. C. 2004, AJ, 127, 1089 

\bibitem[\protect\citeauthoryear{Lynch et al.}{2006}]{Lynch06} Lynch, D. K., et al. 2006, ApJ, 638, 987 

\bibitem[\protect\citeauthoryear{Marshall et al.}{2006}]{Marshall06} Marshall, D. J., Robin, A. C., Reyl\'{e}, C.,  Schultheis, M., Picaud, S. 2006, A\&A, 453, 635

\bibitem[\protect\citeauthoryear{Martini et al.}{1999}]{Martini99} Martini, P., Wagner, R. M., Tomaney, A., Rich, R. M., Della Valle, M., Hauschildt, P. H. 1999, AJ, 118, 1034

\bibitem[\protect\citeauthoryear{Mennella et al.}{1998}]{Mennella98} Mennella, V., Brucato, J. R., Colangeli, L., Palumbo, P., Rotundi, A., Bussoletti, E. 1998, ApJ, 496, 1058

\bibitem[\protect\citeauthoryear{Munari et al.}{1996}]{Munari96} Munari, U., et al. 1996, A\&A, 315, 166 

\bibitem[\protect\citeauthoryear{Munari et al.}{2002}]{Munari02} Munari, U., et al. 2002, A\%A, 389, L51

\bibitem[\protect\citeauthoryear{Pastorello et al.}{2007}]{Pastorello07} Pastorello, A., Della Valle, M., Smarrt, S. J., Zampieri, L., Benetti, S., Cappellaro, E., Patat F., Spiro, S., Turatto, M., Valenti, S. 2007, Nat, 449, 7164

\bibitem[\protect\citeauthoryear{Press et al.}{1992}]{Press92} Press, W., Teukolsky, S., Vetterliung, W., Flannery, B. 1992, Numerical Recipes in Fortran, Cambridge University Press

\bibitem[\protect\citeauthoryear{Rau et al.}{2007}]{Rau07} Rau, A., Kulkarni, S. R., Ofek, E. O., Yan, L. 2007, ApJ, 659, 1536

\bibitem[\protect\citeauthoryear{Rich et al.}{1989}]{Rich89} Rich, R. M., Mould, J., Picard, A., Frogel, J. A., Davies, R. 1989, ApJ, 341, L51 

\bibitem[\protect\citeauthoryear{Rosino et al.}{1992}]{Rosino92} Rosino, L., Iijima, T., Beneti, S., D'Ambrosio, V., di Paolantonio, A., Kolotilov, E. A. 1992, A\&A, 257, 603

\bibitem[\protect\citeauthoryear{Rudy et al.}{2003}]{Rudy03} Rudy, R. J., Dimpfl, W. L., Lynch, D. K., Mazuk, S., Venturini, C. C., Wilson, J. C., Puetter, R. C., Perry, R. B. 2003, ApJ, 596, 1229 

\bibitem[\protect\citeauthoryear{Rushton et al.}{2005}]{Rushton05} Rushton, M. T., Geballe, T. R., Filippenko, A. V., Chornock, R., Li W., Leonard, D. C., Foley, R. J., Evans, A., Smalley, B., van Loon, J. Th., Eyres, S. P. S. 2005, MNRAS, 360, 1281 

\bibitem[\protect\citeauthoryear{Schlegel et al.}{1998}]{Schlegel98} Schlegel, D., Finkbeiner, D., Davis, M. 1998, ApJ, 500, 525.

\bibitem[\protect\citeauthoryear{Souza et al.}{2003}]{Souza03} Souza, W., Aguiar, J. G. S., Pearce, A. 2003, IAUC, 8185

\bibitem[\protect\citeauthoryear{Tabur \& Monard}{2003}]{Tabur03} Tabur, V., Monard, L. A. G. 2003, IAUC, 8134

\bibitem[\protect\citeauthoryear{Warner}{2008}]{Warner08} Warner, B. 2008, in Classical Novae, eds. M. F. Bode \& A. Evans, 2nd edition, Cambridge University Press, in press

\bibitem[\protect\citeauthoryear{Williams et al.}{1996}]{Williams96} Williams, P. M., Longmore, A. J., Geballe, T. R. 1996, MNRAS, 279, 804 

\bibitem[\protect\citeauthoryear{van den Bergh \& Younger}{1987}]{vandenBergh87} van den Bergh, S. Younger, P. F. 1987, A\&AS, 70, 125  

\bibitem[\protect\citeauthoryear{van Hoof}{2006}]{vanhoof06} van Hoof, P. 2006, http://www.pa.uky.edu/$\sim$peter/atomic 

\bibitem[\protect\citeauthoryear{Venturini et al.}{2004}]{Venturini04} Venturini, C. C., Rudy, R. J., Lynch, D. K., Mazuk, S., Puetter, R. C. 2004, AJ, 128, 405 
\end{thebibliography}
\end{document}